\documentclass[preprint]{aastex6}

\usepackage{graphicx}
\usepackage{subcaption}
\definecolor{airforceblue}{rgb}{0.36, 0.54, 0.66}
\hypersetup{citecolor=airforceblue}

\begin{document}

\def\sigmaunit{\ifmmode {\rm\,M_\odot\,yr^{-1}\,kpc^{-2}}\else
                ${\rm\,M_\odot\,yr^{-1}\,kpc^{-2}}$\fi}
\def\kmsMpc{\ifmmode {\rm\,km\,s^{-1}\,Mpc^{-1}}\else
                ${\rm\,km\,s^{-1}\,Mpc^{-1}}$\fi}

\title{correlation between SFR surface density and thermal pressure of ionized gas in local analogs of high-redshift galaxies}

\author{Tianxing Jiang\altaffilmark{1,2,3}, Sangeeta Malhotra\altaffilmark{4}, Huan Yang\altaffilmark{5},
James E. Rhoads\altaffilmark{4}}
\altaffiltext{1}{School of Earth and Space Exploration, Arizona State University, Tempe, AZ 85287, USA; tianxing.jiang@asu.edu}
\altaffiltext{2}{LSSTC Data Science Fellow}
\altaffiltext{3}{Department of Astronomy, University of Maryland, College Park, MD 20742, USA}
\altaffiltext{4}{NASA Goddard Space Flight Center, Greenbelt, MD 20771, USA}
\altaffiltext{5}{Las Campanas Observatory, Carnegie Institution of Washington, La Serena, Chile}

\begin{abstract}
We explore the relation between the star formation rate surface density ($\Sigma$SFR) and the interstellar gas pressure for  nearby compact starburst galaxies. The sample consists of 17 green peas and 19 Lyman break analogs. Green peas are nearby analogs of Ly$\alpha$ emitters at high redshift and Lyman break analogs are nearby analogs of Lyman break galaxies at high redshift. We measure the sizes for green peas using Hubble Space Telescope Cosmic Origins Spectrograph (COS) NUV images with a spatial resolution of $\sim$ 0.05$^{''}$. We estimate the gas thermal pressure in HII regions by $P = N_{total}Tk{_B} \simeq 2n_{e}Tk{_B}$.
The electron density is derived using the [SII] doublet at 6716,6731 \AA\, and the temperature is calculated from the [OIII] lines. The correlation is characterized by $\Sigma$SFR = $2.40\times10^{-3} \sigmaunit (\frac{P / k_{B}}{10^{4}\, {\rm cm}^{-3}\,{\rm K}})^{1.33}$.  Green peas and Lyman break analogs have high $\Sigma$SFR up to 1.2 M$_{\odot\,}$yr$^{-1\,}$kpc$^{-2}$ and high thermal pressure in HII region up to P/k$_B$ $\sim$10$^{7.2}{\rm\, K\, cm}^{-3}$. These values are at the highest end of the range seen in nearby starburst galaxies.
The high gas pressure and the correlation, are in agreement with those found in star-forming galaxies at z $\sim$ 2.5. These extreme pressures are shown to be responsible for driving galactic winds in nearby starbursts. These outflows may be a crucial in enabling Lyman-$\alpha$ and Lyman-continuum to escape.

\end{abstract}


\section{Introduction}

Understanding the physical factors that control or affect star formation in galaxies is one of the most critical aspects of understanding galaxy evolution. Star formation is linked to the interstellar medium. On galactic scales, cold clouds collapse under its own gravity, fragment into small dense cores, and eventually stars form there. Stars inject energy, momentum, metals and gas into the interstellar medium by stellar feedback (e.g. stellar winds, radiation, and supernova explosion), and ionize and heat the interstellar medium. Hot, ionized gas then cools and converts to cold gas again. Empirical star formation scaling relations are essential input for models and simulations of galaxy evolution \citep[e.g.][]{spr03}, due to the complexity of star formation physics.

Observationally, on galactic scales, the star formation rate surface density ($\Sigma$SFR) in galaxies correlates with the neutral gas (atomic and molecular gas) surface density by the empirical ``Kennicutt-Schmidt law" \citep[e.g.][]{sch59,ken89,ken98}. This correlation has also been investigated on sub-galactic scales \citep[e.g.][]{won02,bli04,bli06,big08,roy15}. $\Sigma$SFR is also proposed to be related to the galactic orbital time $\Omega$ \citep[e.g.][]{ken98,won02,gen10,dad10,gar12}, or to the stellar mass surface density \citep[e.g.][]{boi03,shi11,rah16}. However, these relations are often more complex than a simple mathematical expression and can vary in different types of galaxies. How the star formation in galaxies is controlled and regulated is still not quite clear. Based on numerical simulations of multiphase gaseous disks, \citet{kim11} discussed the relation between $\Sigma$SFR and the total midplane pressure of diffuse interstellar medium for star-forming disk galaxies in the regime where diffuse atomic gas dominates the interstellar medium \citep[see also][]{ost11}. Among many physical properties they explored using numerical simulations, the best star formation correlation  they have found is with the total midplane pressure of diffuse interstellar medium. They argued that this correlation should also apply to the starburst regime (generally where gas density $\Sigma$ $\sim$ $10^{2}$ $-$ $10^{4}$ $M_{\odot}pc^{-2}$), such as (ultra) luminous infrared galaxies ([U]LIRGs) and galactic centers. 

The question naturally arises of what the observations tell us about the potential relation between the star formation and the gas pressure in galaxies. Is there a good correlation? One way to measure the pressure is from the gas density and gas temperature. For ionized gas, the thermal pressure $P = N_{total}Tk{_B} \simeq 2n_{e}Tk{_B}$, where the  electron density $n_{e}$ is not hard to measure with more and more available high-quality high-resolution rest-frame optical spectra for both z $\sim$ 0 and z $\sim$ 2 galaxies \citep[e.g.][]{hai09,leh09,ste14,san16}. 

Two studies indirectly suggest the association of star formation rate with the electron density in star-forming galaxies. This might also suggest the association of star formation rate with the thermal pressure of ionized gas, with the assumption that the temperature of ionized gas is comparable in these galaxies. \citet{liu08} showed histograms of the specific star formation rate (SFR/M$_{\ast}$ or sSFR), SFR surface density ($\Sigma$SFR), and [SII]$\lambda$6716/[SII]$\lambda$6731 ratio for SDSS Main sample (typical star-forming galaxies) and SDSS Offset-SF sample galaxies in their Fig.10. They have reported that the Offset-SF sample have both higher $\Sigma$SFR and higher electron density (thus higher pressure in HII regions) compared to SDSS Main sample. It was claimed that the higher SFR surface density may account for the higher interstellar pressure seen in the HII regions of Offset-SF objects. \citet{bri08} investigated the trends of SFR/M$_{\ast}$, $\Sigma$SFR, and [SII]$\lambda$6716/[SII]$\lambda$6731 ratio with their position in the [OIII]$\lambda$5007/H$\beta$ vs [NII]$\lambda$6583/H$\alpha$ BPT diagram for SDSS galaxies. They have found that the galaxies more away from the mean SDSS star-forming abundance sequence are characterized by higher SFR/M$_{\ast}$, $\Sigma$SFR and higher electron density. Neither studies directly presented the relation between $\Sigma$SFR and electron density. \citet{shi15} directly showed the correlation between $\Sigma$SFR and the electron density $n_{e}$ and the correlation between the sSFR and $n_{e}$ for star-forming galaxies at z $\sim$ 2.5, with a sample of 14 H$\alpha$ emitters. \citet{san16} found no correlation between sSFR and $n_{e}$ using a larger sample at z $\sim$ 2.3, but they did not investigate the correlation between $\Sigma$SFR and $n_{e}$. \citet{bia16} studied the median electron density in different sSFR and $\Sigma$SFR bins. They have found that for typical SDSS star-forming galaxies, for a fixed sSFR, the electron density increases with increasing $\Sigma$SFR, but for a fixed $\Sigma$SFR, the electron density deceases with increasing sSFR. This trend was not found for their ``local analogs". \citet{her17} have found that the thermal pressure of the diffuse neutral gas increases with $\Sigma$SFR in nearby galaxies.

In this work, we look into the relation between the SFR surface density $\Sigma$SFR and the interstellar gas pressure on galactic scales. We seek to add observational constraints to the theories and simulations of the interplay between star formation and interstellar medium on galactic scales in the context of galaxy evolution. We study quantitatively the relation between $\Sigma$SFR and thermal pressure of ionized gas for nearby compact starburst galaxies, with the sample of green peas and Lyman break analogs. Green peas are nearby analogs of high-redshift Ly$\alpha$ emitters \citep[e.g.][]{jas14, hen15, yang16, yang17b, ver17}. Lyman break analogs are the counterparts in the nearby universe of the high-redshift Lyman break galaxies (LBGs) \citep{hec05}. Both of them provide best local laboratories for us to study the physical properties of the high-redshift star-forming galaxies, which is why we are particularly interested in these galaxies. We would like to see if there is a $\Sigma$SFR - P$_{gas}$ correlation for these galaxies, and if so, how it compares with that for z $\sim$ 2.5 galaxies. We adopt the cosmological parameters of $\Omega_{M}$=0.3, $\Omega_{\Lambda}$=0.7 and $H_{0}=70\kmsMpc$ throughout this paper.

\section{Data Sample}

Green pea galaxies were first noted by volunteers in the Galaxy Zoo project \citep{lin08}. They looked green and appeared to be unresolved round point sources in the gri composite color image \citep{car09} from the Sloan Digital Sky Survey \citep[SDSS]{yor00}. Our sample of green peas is taken from the catalog in \citet{car09}. By defining a color selection in the redshift range 0.112 $\leq$ z $\leq$ 0.360, \citet{car09} systematically selected 251 green peas with extreme [OIII]$\lambda$5007 equivalent widths from the SDSS Data Release 7 (DR7) spectroscopic data base.  80 out of 251 are star-forming objects that have high S/N SDSS spectra. These star-forming green peas are low-mass galaxies with high star formation rates and low metallicity. For these 80 star-forming green peas, 12 of them have NUV (near-UV) images taken with the Cosmic Origins Spectrograph (COS) in HST archive (PIs: Henry (GO: 12928); Jaskot (GO: 13293); Heckman (GO: 11727)) and were discussed in \citet{hen15, yang16, yang17a}, and 19 of them have COS NUV images from our recent HST observation (PI: Malhotra (GO: 14201)). To get a well-measured size of the galaxies, the galaxies have to be spatially resolved. We emphasize that these COS NUV images offer a tremendous gain in resolution (of $\sim$ 0.05$^{''}$)  over that of SDSS images (PSF width $\sim$1.4$^{''}$). The seeing of SDSS images is larger than the SDSS r-band half-light radii of green peas.

Lyman break analogs (LBAs) are supercompact UV luminous galaxies originally selected by \citet{hec05} as local starburst galaxies that share typical characteristics of high-redshift LBGs. They are star-forming galaxies at $z < 0.3$ that satisfy the criteria L$_{FUV} > 10^{10.3} L_{\odot}$ and I$_{FUV} > 10^{9}L_{\odot}\, \hbox{kpc}^{-2}$. LBAs share similar stellar mass, metallicty, dust extinction, SFR, physical size and gas velocity dispersion with Lyman break galaxies. Our sample of Lyman break analogs is drawn from \citet{ove09}. We excluded 6 out of 31 LBAs as these 6 objects have dominant central objects and might be Type 2 AGNs. We used the optical half-light radius from their Table 1. The radii are either from HST WFPC2 F606W images (PSF FWHM $\sim$ 0.11$^{''}$) or from HST ACS Wide Field Channel F850LP images (PSF FWHM $\sim$ 0.12$^{''}$). 

There are optical spectra in SDSS Data Release 12 (DR12) spectroscopic data base with well-resolved [SII]$\lambda$$\lambda$6716,6731 lines \citep{ala15} for the 31 green peas and for 24 LBAs out of the 25 LBAs. With visual inspection of the spectra, we excluded two green peas and two LBAs as the [SII]$\lambda$$\lambda$6716,6731 lines in SDSS spectra are badly contaminated by the sky lines. One of the green peas was also included as a Lyman break analog in \citet{ove09}. We include this one in the sample of Lyman break analogs in our work and do not count it twice.  Of the remaining 50 objects, all but 3 have emission line measurements and SFR measurements in the public MPA-JHU catalogs\footnote{Available at data.sdss3.org/sas/dr8/common/sdss-spectro/redux/}, which are based on SDSS Data Release 8 (DR8). In total, we end up with 47 objects, 26 green peas and 21 LBAs. We refer to them as the ``parent sample''. 
We decided to use MPA-JHU catalogs in our work instead of the pipeline measurements from SDSS DR12 for two primary reasons. First, the emission line fluxes are better measured in MPA-JHU measurements by using stellar population synthesis models to accurately fit and subtract the stellar continuum; while for SDSS pipeline measurements, the emission line fluxes are measured by fitting multiple Gaussian-plus-background models to the lines. We can get more accurate [SII] measurements as needed. Second,the total SFR (using the galaxy photometry as described in \citet{sal07}) and fiber SFR (using H$\alpha$ fluxes within the galaxy fiber aperture as described in \citet{bri04}) are provided by MPA-JHU measurement. 

We have derived our own star formation rates independently (see section~3.3) but take advantage of the information in the MPA-JHU catalog to correct for the extended light outside the fiber as part of our procedure.

\section{Method}

\subsection{Electron Density}

The average electron density in a nebula can be measured by observing the effects of collisional de-excitation. This can be done by comparing the intensities of two lines of a single species emitted by different levels with nearly the same excitation energy and different radiative transition probabilities or different collisional de-excitation rates  \citep[see, e.g., Chapter 5 of][]{ost2006book}. The ratio of the intensities of the lines they emit depends on the relative populations of the two levels, which is dependent on the collision strengths of the two levels. So the ratio of the intensities of the lines is sensitive to the electron density. The most frequently used emission line doublets in rest-frame optical spectra are [OII]$\lambda$$\lambda$3726,3729 and [SII]$\lambda$$\lambda$6716,6731. Since the SDSS spectra do not properly resolve [OII]$\lambda$$\lambda$3726,3729 but do resolve [SII]$\lambda$$\lambda$6716,6731, we measured the electron density from [SII] doublets. The [SII] doublet ratio is a good measurement of the electron density for $10^{1.5} cm^{-3} < n_{e} < 10^{3.5} cm^{-3}$. The program ``temden'' under the IRAF STS package NEBULAR is available for the measurement with input of the intensity ratio of the doublets and temperature. The output electron density is insensitive to the input temperature for $7500 K < T_{e} < 15000 K$. When measuring Ne, we assumed $T_{e}= 10{^4}$K, which is an order-of-magnitude estimate for HII regions. \citet{san16} have argued that the measurement of the electron density is different when using the most up-to-date collision strength and transition probability atomic data instead of the old values included in the IRAF routine temden. However, we notice that the measurements of $n_{e}$ from [SII] doublets based on either the updated value in \citet{san16} or IRAF temden are very close to each other for $10^{1.5}  cm^{-3} < n_{e} < 10^{3.5} cm^{-3}$ , with differences of n$_e$ at a fixed [SII] ratio within $\sim$0.1 dex, as seen in Fig.1 in \citet{san16}. 
 
The line ratio is R = $\frac{[SII]\lambda6716}{[SII]\lambda6731}$. The lower uncertainty and upper uncertainty of the ratio are calculated separately: the lower uncertainty is l$_{err}$= R - $\frac{[SII]\lambda6716 - [SII]\lambda6716_{err}}{[SII]\lambda6731 + [SII]\lambda6731_{err}}$, the upper uncertainty is u$_{err}$  = $\frac{[SII]\lambda6716 + [SII]\lambda6716_{err}}{[SII]\lambda6731 - [SII]\lambda6731_{err}} - R$. We only measured the electron density for the objects that have more than 4$\sigma$ detection of [SII]$\lambda$6716 and [SII]$\lambda$6731 and satisfy $\frac{R}{l_{err}} > $3 and $\frac{R}{u_{err}} > $3 (38 objects out of 47 objects in the ``parent sample''). As seen from the dashed line in Fig.1, in both very high (with ratio lower than $\sim$ 0.44) and very low electron density regime (with ratio higher than $\sim$ 1.38), the line ratio is not sensitive to the electron density at all. And the theoretical maximum of the line ratio is $\sim$ 1.43. Taking these into account, we classify the measurement of the electron density into four cases. 1. If the lower bound of the line ratio is higher than 1.38, we can only measure the upper limit of electron density, which corresponds to the line ratio of 1.38. 2. If the lower bound of the line ratio is between 1.10 and 1.38 and the upper bound of the line ratio is higher than 1.38,  we can only measure the upper limit of electron density, which corresponds to the lower bound of the line ratio. 3. If the lower bound of the line ratio is less than 1.15 and the upper bound of the line ratio is higher than 1.38, the uncertainty of the electron density spans a wide range and thus the measurement is not useful. 4. If the upper bound of the ratio is not higher than 1.38, then we can safely measure the electron density and its (upper and lower) uncertainty. For the fourth case, the lower (upper) uncertainty of the electron density corresponds to the upper (lower) uncertainty of the line ratio. We throw away 2 objects that are classified in the third case. Therefore, there are 36 objects that have electron density measurements out of the 47 objects in the ``parent sample''.

Fig. 1 shows the line ratios and electron density measurements based on the IRAF ``temden'' package for the remaining 36 objects out of the ``parent sample''. There are 17 green peas and 19 LBAs in Fig.1. We call them the ``final sample''. Note that in Fig.4, the thermal pressure is only measured for the ``final sample''. And in Table 1, the properties are also for the ``final sample'' instead of the``parent sample''. 

The dashed line in Fig. 1 is the fitted function $R(n_{e}) = a\frac{b + n_{e}}{c + n_{e}}$ between $n{_e}$ and the line ratio R over a range of electron densities of 10cm$^{-3}$ to 10$^4$cm$^{-3}$ for the temden package, similar to what has been done in \citet{san16}. The result is $R(n_{e}) = a\frac{b + n_{e}}{c + n_{e}}$, with a = 0.4441, b = 2514, and c  = 779.3. 

As seen from Fig.1, the electron densities for our ``final sample'' are mostly 100 $\sim$ 700 cm$^{-3}$. This is comparable to the typical electron densities for z $\sim$ 2 star-forming galaxies \citep{ste14, shi15, san16, kas17} and much larger than the typical electron densities ($\sim$30 $cm^{-3}$ or 10 -- 100 $cm^{-3}$ ) measured for SDSS star-forming galaxies \citep{bri08,san16}.

\subsection{Electron Temperature}

The electron temperature in a nebula can be determined from measuring the ratio of intensities of two lines of a single species emitted from two levels with considerably different excitation energies \citep[Chapter 5 of][]{ost2006book}. In rest-frame optical spectra, the most frequently used emission lines are [OIII]$\lambda$5007,[OIII]$\lambda$4959 and [OIII]$\lambda$4363. Since these three lines are relatively close in wavelength, the effect of dust extinction on the ratio of $\frac{[OIII]\lambda5007 + \lambda4959}{[OIII]\lambda4363}$ is small. In the ``parent sample'' of 47 objects, 36 objects have at least 2$\sigma$ detection of [OIII]$\lambda$5007, [OIII]$\lambda$4959 and [OIII]$\lambda$4363. For these 36 objects, the ratio of  $\frac{[OIII]\lambda5007 + \lambda4959}{[OIII]\lambda4363}$ was input to the program ``temden'' in IRAF to measure the temperature. Therefore, in the ``parent sample'' of 47 objects, 36 objects have electron temperature measurements. For these 36 objects, the typical uncertainties are 200 - 1500 K and the median uncertainty is -497K, +612K. Among the ``final sample'' of 36 objects that have electron density measurements from section 3.1, only 26 of them have electron temperature measurements. For the other 10 objects in the ``final sample", we assumed a temperature of 11000 K. Among the 10 objects, there are two objects with at least 2$\sigma$ detection of [OIII]$\lambda$5007, [OIII]$\lambda$4959 and S/N of [OIII]$\lambda$4363 between 1.5 and 2 in our ``final sample'', for which the electron temperature is 11300$_{-1490}^{+4440}$K and 11400$_{-1420K}
^{+3740K}$. The assumed 11000K for these 10 objects in our ``final sample'' is consistent with the temperature of these two objects, and with the uncertainties or the lower limits on the line ratios of these 10 objects. The assumed 11000 K is also close to the median temperature of 12391 K (11$\%$ difference) of the 36 objects in the ``parent sample'' but slightly lower, as befits a subset of objects with somewhat weaker [OIII]$\lambda$4363 emission.

Fig.2 shows the distribution of the electron temperatures for 36 objects out of the ``parent sample''. The electron temperature is mostly 10000 K - 15000 K. \citet{and13} measured electron temperature from O$^{++}$ for stellar mass-SFR stacks of SDSS galaxies, which is mostly between 10500 K and 12000 K. In comparison, the electron temperature of our sample is slightly larger than the typical electron temperature in z $\sim$ 0 SDSS star-forming galaxies.

\subsection{Star Formation Rate}

We measured the SFRs from the H$\alpha$ fluxes in MPA-JHU catalogs. The line fluxes from MPA-JHU catalogs have been corrected for Galactic extinction following \citet{odo94} attenuation curve. First we derived dust extinction in the emitting galaxy assuming the \citet{cal00} extinction curve and an intrinsic H$\alpha$/H$\beta$ value of 2.86: $E (B - V)_{gas} = \frac{\log_{10}[(f_{H\alpha}/f_{H\beta})/2.86]}{ 0.4\times[k(H\beta) - k(H\alpha)]}$, $A_{H\alpha} = k(H\alpha)E(B - V)_{gas}$, with k(H$\alpha$) = 2.468 and  k(H$\beta$) - k(H$\alpha$) = 1.163. Then the SFR was calculated by SFR ($M_{\odot}yr^{-1}$) = $10^{-41.27}L_{H\alpha,corr}$ (erg s$^{-1}$) according to \citet{ken12}. That is our own fiber SFR. The SFRs are not sensitive to the dust extinction law chosen, because the dust extinction is low ((B - V)$_{gas}$ $\sim$ 0.1 mag) for our sample. The SFR will change no more than 0.03 dex if the extinction law from the Milky Way (MW) the Large Magellanic Clouds (LMC), or the Small Magellanic Clouds (SMC) is chosen instead of the \citet{cal00} extinction. We calculated the ratio of the total SFR to the fiber SFR that are both available in MPA-JHU catalogs. For green peas the ratios are typically less than 1.2, and for LBAs typically around 1.5. Then we corrected our own fiber SFR by applying the factor of this ratio. For LBAs, we compared the SFR based on MPA-JHU with the SFR measurements from  H$\alpha$ luminosity in \citet{ove09}. Note that \citet{ove09} applied a small correction factor to H$\alpha$ fluxes of typically $\sim$ 1.7 due to the flux expected outside the SDSS fiber. We found good statistical agreement and no gross systematic differences between the SFR based on MPA-JHU and the SFR in \citet{ove09}.

\subsection{Half-light Radius}

GALFIT \footnote{http://users.obs.carnegiescience.edu/peng/work/galfit/galfit.html} is an image analysis algorithm that can model the light distribution of galaxies, stars, and other astronomical objects in 2 dimensional digital images by using analytic functions. We measured the half-light radii of the green peas from COS NUV images using GALFIT version 3.0 \citep{pen10}. The Sersic radial profile, which is one of the most frequently used profiles for galaxy morphology analysis, was chosen in our measurement. The distribution of the UV half-light radii for green peas is shown in Fig. 3. The typical radii is $\sim$0.19 arcsec, and $\sim$0.7 kpc, as listed in Table 1.
  
To estimate the UV sizes of Lyman break analogs, the optical sizes of Lyman break analogs were divided by a representative value of 1.8, considering that the optical size is typically (about 2 times) larger than the UV size for Lyman break analogs \citep{ove08}. We do not apply PSF image in GALFIT for the size and sersic index measurement. The effects of PSF should be small, as the sizes we measured are more than 3 times bigger than the PSF FWHM, with only three exceptions whose sizes were overestimated by up to $\sim$ 10$\%$.  

\section{Results}

For the 36 objects in the ``final sample'', we measured the thermal pressure in the HII region by $P/k{_B} = N_{total}T$. If helium is singly ionized, then $N_{total} \simeq n_e + n_{H^+} + n_{He^+} \simeq 2n_e$. If some helium is doubly ionized, then the N$_{total}$ could be slightly less than 2$n_e$. Since the number density of helium atom+ion is only around 8$\%$ of the H$^+$ density, this should be a minor effect. the ionization potential of Sulfur is 10.36 eV, lower than the ionization potential of Hydrogen. So [SII] doublets also exist beyond the boundary of HII regions, where there are neutral hydrogen atoms in addition to the electrons and protons. So N$_{total} = 2n_{e}$ is a lower limit of the total ion and atom density. We also calculated the $\Sigma$SFR by $\Sigma$SFR  = $\frac{SFR/2}{\pi \times R_{e}^{2}}$. 

The thermal pressure in HII regions and the $\Sigma$SFR are shown in Fig.4. We have included the uncertainties of the electron density and the temperature in the pressure uncertainty for each object. Note that for the 10 objects with an assumed temperature of 11000K, we took -1460K, +4090K (the average of  -1490K, +4440K, and  -1420K , +3740K) as representative uncertainties of the temperature. We find that our local analogs have high $\Sigma$SFR up to 1.2 $M_{\odot} {\rm yr^{-1}\, kpc^{-2}}$ and high thermal pressure in HII region up to P/k$_B$ $\sim$10$^{7.2}{\rm K\,cm}^{-3}$. 

The thermal pressure of our sample is higher than that for typical SDSS star-forming galaxies with thermal pressure around P/k$_B$ = 10$^{5.8}$ Kcm$^{-3}$ (when n$_e$ = 30 cm$^{-3}$ and T = 11000 K are taken). In addition, green peas have higher average $\Sigma$SFR and higher average thermal pressure than Lyman break analogs. The thermal pressures seen in green peas are near the upper end of pressures seen in starbursts by \citet{hec90}. In nearby starbursts, these extreme pressures are responsible for driving galactic outflows (\citet{hec90}), which are necessary for the resonantly scattered Lyman-$\alpha$ photons to escape.

To quantitatively describe the correlation, we used Spearman's rank correlation, a non-parametric test for correlation. Spearman's correlation coefficient $r_{s}$ measures the strength of association between two ranked variables. And the corresponding $p$-value tells you significance level with which a null hypothesis that the variables are unrelated can be rejected. Spearman's rank correlation does not handle upper limits or error bars, so for the objects that only have upper limits for the electron density, we ``re-measured'' their electron density only for the purpose of applying Spearman's rank correlation. For the objects with R $>$ 1.5, we could not get a reliable electron density measurement, so we excluded them from the Spearman's rank correlation analysis. For the objects that with R $\le$ 1.38, we measured the electron density from the line ratio (without considering the error bars). For objects with 1.38 $<$ R $<$ 1.5, we measured the electron density from a ratio of 1.38. See the column ``n$_{e}$" in Table 1 for the measurements of the electron density that are used for Spearman's rank correlation. Then we measured the pressure again combining the new electron density measurements here and the temperature measurements from section 3.2. This is shown in Fig.5.

We calculated $r_{s}$ and p-value for the data points in Fig.5, and obtained $r_{s}$ = 0.615 and p = 0.02$\%$. We checked that if we did not apply the correction factor (for the extended light outside of the fiber) to the SFR, we would obtain $r_{s}$ = 0.598 and p = 0.05$\%$ and we would still see the correlation.

The next step is to fit a linear function between log$\Sigma$SFR and log(P/K$_{B}$), where P/K$_{B}$ denotes the thermal pressure. Since the relation between [SII] line ratio and electron density is non-linear, it is harder to know the distribution of the uncertainties of the electron density (obviously it is not appropriate to assume that the distribution of the uncertainties is close to gaussian), and thus the distribution of the uncertainties of the thermal pressure. Moreover, it is hard to deal with the upper limits of the thermal pressure if fitting directly to log $\Sigma$SFR and log P/K$_{B}$. Instead, we did a 2-dimensional fitting to the [SII] line ratio, the electron temperature and log($\Sigma$SFR).

We assumed a linear relation between log P/K$_{B}$ and log$\Sigma$SFR,  
$$log P/k_{B} = f \times log\Sigma SFR + g  , $$ 
where f and g are two unknown parameters. Then 
$$log(2n_{e}T) = f \times log(\Sigma SFR) + g ,$$ 
$$n_{e}(\Sigma SFR,T) = \frac{10^{(log (\Sigma SFR ^ f ))
}   \times 10^{g}}{2 \times T} .$$ 
Plugging this into $R(n_{e}) = a\frac{b + n_{e}}{c + n_{e}}$, we know 

$$R(\Sigma SFR,T)   =  a\times \frac{b + \frac{10^{(log (\Sigma SFR ^ f ))
}  \times 10^{g}}{2 \times T}}{c + \frac{10^{(log (\Sigma SFR ^ f))
}   \times 10^{g}}{2  \times T}} .$$

We took the function R($\Sigma$SFR,T) in the 2-dimensional fitting, to figure out the values of parameters f and g for the best-fit. Note that the uncertainty of the temperature and the uncertainty of $\Sigma$SFR are small, compared to the uncertainty of R = $\frac{[SII]\lambda6716}{[SII]\lambda6731}$. We applied weighted least-squares fitting to this 2-dimensional fitting. This is only valid when the uncertainties of the line ratio R are gaussian. But it should not be a bad assumption to take the uncertainties of the ratio as approximately gaussian just for a rough estimate of the parameters f and g. Since the lower and upper uncertainties of the ratio are not symmetric, we used the larger one for each pair of lower and upper uncertainties in the weighted least-square fitting. The parameters f, g for the best fit of R($\Sigma$SFR,T) are 0.750, 5.966, respectively. So the best fit in terms of log (P/k$_{B}$) and log $\Sigma$SFR is
$$log(P/k_{B}) = 0.750 \times \log\Sigma SFR + 5.966. $$
This can be rewritten as 
$$\Sigma SFR = 10^{-7.95}M _{\odot}yr^{-1}kpc^{-2} \times(P/k_{B})^{1.33}, $$
or 
$$\Sigma SFR = 2.40\times 10^{-3}M_{\odot}yr^{-1}kpc^{-2}\left(\frac{P/k_{B}}{10^{4}cm^{-3}K}\right)^{1.33}.$$

The best-fit exponent is 1.33, and the 68$\%$ confidence interval of this exponent is 1.08 -- 1.74. The best fit is shown in Fig.4 as the purple line.

For the subset of data points that have 1$\sigma$ uncertainties on the pressure (instead of upper limits) in Fig.4, the scatter (1$\sigma$ standard deviation) of the pressure around the best fit is 0.268 dex, while the median pressure measurement uncertainty for this subset is -0.300 dex, +0.248 dex. So the scatter is mostly due to the measurement uncertainties.

\section{Discussion}

\subsection{Contribution from Diffuse Ionized Gas}

In our work, we are interested in the pressure and the electron density inside HII regions. However, the [SII] fluxes we measured are from the spectra of the whole galaxy, including HII regions (and beyond the boundary of HII regions) and the diffuse warm ionized gas. Therefore, the estimated electron density based on the integrated-light galaxy spectra may not well represent the real electron density of HII regions. We treat the emission from diffuse ionized gas as contamination to [SII] fluxes in this work. It is hard to know exactly the effects of contamination from the diffuse ionized gas. Here we provide a rough estimate of the effects of [SII] fluxes from the diffuse ionized gas on the measurement of the electron density of the HII regions, based on the (unrealistic) assumption that there are purely two components emitting [SII] in the galaxy, each with a uniform electron density. The estimate here should be treated as a toy model. There are some work studying the properties of the diffuse ionized gas in different galaxies, such as, \citet{haf99} and \citet{mad06} using the Galaxy,  \citet{hid07} using the dwarf irregular galaxy NGC 6822, \citet{flo09} using a set of 29 galaxies from the literature including 25 spirals and 4 irregulars, and \citet{mon10} using luminous and ultraluminous infrared galaxies. $[SII]\lambda6716/H\alpha$ is higher in diffuse ionized gas compared to HII regions. For the Galaxy, $[SII]\lambda6716/H\alpha$ in diffuse ionized gas and in HII regions is around 0.38 and 0.12 \citep{mad06}, respectively. The difference of $[SII]\lambda6716/H\alpha$ in diffuse ionized gas and HII regions is smaller in the dwarf irregular galaxy \citep{hid07} than in the Galaxy. We took $\frac{[SII]\lambda6716}{H\alpha}  = 0.125$ for diffuse ionized gas and $\frac{[SII]\lambda6716}{H\alpha}  = $ 0.090 for HII regions as the representative values for our sample from the dwarf irregular galaxy NGC 6822 \citep{hid07} and take $\frac{[SII]\lambda6716}{H\alpha}  = 0.38$ for diffuse ionized gas and $\frac{[SII]\lambda6716}{H\alpha}  = 0.12$ for HII regions as the representative values for star-forming spirals. In addition, we have assumed that the ratio of H$\alpha$ luminosity coming from HII region and diffuse ionized gas is 5:5 for spirals (Sb and Sc) and that the ratio is 7: 3 for our sample (starbursts) \citep[Fig.8 in ][]{oey07}. In our estimate, we took three different values for the electron density of diffuse ionized gas: $n_{e,DIG}$  = 0.5 $cm^{-3}$, 10 $cm^{-3}$ and 50 $cm^{-3}$. Recall that we fitted a function $R(n_{e}) = a\frac{b + n_{e}}{c + n_{e}}$, so the theoretical line ratio in diffuse ionized gas (DIG) is $R_{DIG} = a\frac{b + n_{e,DIG}}{c + n_{e,DIG}}$. 

For dwarf irregular starbursts, 
\[
R_{observed} =  \frac{L_{6716,DIG} + L_{6716,HII}}{L_{6731,DIG} + L_{6731,HII}}  =   \frac{0.125 \times L(H\alpha, DIG) + 0.090 \times L(H\alpha, HII)}{\frac{0.125 \times L(H\alpha, DIG)}{R_{DIG}} + \frac{0.090 \times L(H\alpha, HII)}{R_{HII}}} ,
\]
so 
\[
R_{observed}  =  \frac{0.125 \times 0.3 + 0.090 \times 0.7}{\frac{0.125 \times 0.3}{R_{DIG}} + \frac{0.090 \times 0.7}{R_{HII}}} ,
\]
where L stands for luminosity. That is, 
$$R_{HII} = \frac{0.090 \times 0.7}{\frac{0.125 \times 0.3 + 0.090 \times 0.7}{R_{observed}} - \frac{0.125 \times 0.3}{R_{DIG}}} ,$$
where $R_{HII}$ is the ratio of the fluxes of [SII] doublets that are emitted from HII regions. From the relation $R(n_{e}) = a\frac{b + n_{e}}{c + n_{e}}$, we know that the real electron density in HII regions is $n_{e,HII} =  \frac{(c\times R_{HII}-a\times b)}{(a-R_{HII})}$. So $n_{e,HII}$ can be written as a function of $R_{observed}$ and $R_{DIG}$, and thus a function of $R_{observed}$ and $n_{e,DIG}$. For spiral galaxies, the demonstration process is the same. We compare the real electron density in HII region and the electron density measured directly from the integrated luminosity in Fig.6. The left panels are for spiral galaxies, and the right panels are for dwarf irregular starbursts. According to Fig.6, for irregular dwarf starbursts (representative of our sample) the electron density in HII region is underestimated by $\sim$ 0.2 -- 0.4 dex, for spirals it is underestimated by $\sim$ 1.0 dex. For irregular dwarf starbursts, the three different assumptions of the electron density in DIG give roughly the same result, while for spirals this assumption matters when the measured electron density from integrated luminosity is lower than $10^{2.5}$ cm$^{-3}$. We argue that we are not sure whether all the objects in our sample resemble the cases of a dwarf irregular starburst galaxy in the left panels of Fig.6, so we show the cases of star-forming spirals as well, as an extreme limit.

One way to get a good measurement of the electron density in HII regions is to use Integrated Field Unit (IFU) measurements or use other line pairs that mainly originate from HII regions and are sensitive to 10$^{2}$cm$^{-3}$ $<$ n$_{e}$ $<$ 10$^{4}$cm$^{-3}$, such as [OIII] 88/52 $\mu$m, [SIII] 33/19 $\mu$m in the infrared. In addition, we should note that the emission lines used for the electron density and electron temperature measurements for the whole galaxy is surface-brightness-weighted. Even inside the HII region or among different HII regions the electron density and the electron temperature can present a gradient. Integrated Field Unit (IFU) measurement can help with this issue.

\subsection{Diffuse Gas as a Possible Cause for Correlation?}
Is it possible that the lower pressure in HII regions of lower $\Sigma$SFR galaxies is due to varying contribution of DIG in low SFR surface density galaxies and high SFR surface density galaxies? Below we discuss the possible different ``extent of underestimate'' of HII region pressure in galaxies with different $\Sigma$SFR.

If lower $\Sigma$SFR galaxies have a higher fraction of [SII]$\lambda$$\lambda$6716,6731 emission coming from DIG than high $\Sigma$SFR galaxies, then lower $\Sigma$SFR galaxies will suffer a more substantial underestimate of the electron densities and pressure in HII regions. How should we compare this fraction in low $\Sigma$SFR galaxies and high $\Sigma$SFR galaxies? In the extreme case when all these galaxies have nearly the same [SII]$\lambda$$\lambda$6716,6731/H$\alpha$ in DIG, and nearly the same [SII]$\lambda$$\lambda$6716,6731/H$\alpha$ in HII region, the observed [SII]$\lambda$$\lambda$6716,6731/H$\alpha$ normalized by metallicity for these objects should directly imply the fraction of [SII]$\lambda$$\lambda$6716,6731 emission coming from DIG (the higher [SII]$\lambda$$\lambda$6716,6731/H$\alpha$ is, the higher the fraction of [SII]$\lambda$$\lambda$6716,6731 emission coming from DIG is). We measured direct T$_e$-based metallicities for 19 objects out of the ``final sample'' (Jiang et al. in preparation). We find that there is no prominent anti-correlation between $\Sigma$SFR and observed [SII]$\lambda$$\lambda$6716,6731/H$\alpha$ normalized by metallicity. However, given that starburst galaxies usually have a small fraction of DIG \citep{cal19, oey07}, we consider it unlikely that the whole trend in Fig.4 is driven by differential contribution of DIG in different galaxies.

\subsection{Comparison with Correlation at High Redshift}

Our study observationally indicates that the nearby compact starburst galaxies with higher SFR surface density tend to have higher thermal pressure in HII regions. 

\citet{shi15} presented the relation between electron density and $\Sigma$SFR for the H$\alpha$ emitters at z $\sim$ 2.5. Note that [OII]$\lambda$$\lambda$3726,3729 are used as tracers of the electron density in \citet{shi15}, while [SII] doublets are used in our work. We estimate the HII region thermal pressure for their sample using $P = 2n_{e}Tk_{B}$, where we assume T  = $10^{4}$K. We compare these galaxies at z $\sim$ 2.5 to our sample. As shown in Fig.4, the H$\alpha$ emitters at z $\sim$ 2.5 obey very similar $\Sigma$SFR correlation with thermal pressure in HII regions to our starburst galaxies at z $<$ 0.3. Note that our sample is larger than the sample in \citet{shi15}. For the same $\Sigma$SFR, the thermal pressure in HII regions in z $\sim$ 2.5 galaxies is comparable to that in local (z $<$ 0.3) analogs (green peas and LBAs). Since green peas and Lyman break analogs are best analogs of high-redshift Ly$\alpha$ emitters and high-redshift Lyman break galaxies, the high-redshift Ly$\alpha$ emitters and high-redshift Lyman break galaxies might also have a similar correlation.

\subsection{Interpretations of the Correlation}

There could be different physical causes for the correlation between SFR surface density and thermal pressure in HII regions. We discuss them as follows.

1. As HII regions evolve, they expand because they are overpressured, and the HII region thermal pressure could drop. The ionizing photon rate due to the UV fluxes of massive stars also drops after around 5 Myr after the burst, thus the H$\alpha$ luminosity drops. This could play a role in the correlation observed in this work. We have measured the ages of the young starbursts in 19 objects out of the ``final sample'' by performing SED fitting to binned SDSS spectra (Jiang et al. in preparation). We do not find systematically older starburst ages among the galaxies having lower SFR surface density and lower thermal pressures. Therefore, this scenario should not be the primary cause of the observed correlation for local analogs. In fact the UV emission from the green peas in our sample is dominated by very young populations (mean age of 5-6 Myr).
 
2. The positive-correlation found in section 4 between $\Sigma$SFR and thermal pressure in HII regions is expected if the thermal pressure is mainly driven by stellar feedback. For example, the mechanical energy injection due to stellar winds and/or supernovae in star-forming regions can increase the gas pressure \citep{str}. \citet{hec90} show that in case of starbursts with strong galactic outflows the pressure is dominated by thermal pressure.

\subsection{Comparison with the Simulation Work}
From the literature we found simulation work by \citet{kim11} that reported a correlation between $\Sigma$SFR and gas pressure. It is interesting to compare it with this work. \citet{kim11} conducted numerical simulations of multiphase gaseous disks in the diffuse-atomic-gas-dominated regime ($\Sigma = 3 - 20 M_{\odot}pc^{-2}$). The simulations span a few hundred Myr, and the disks  evolve to a state of vertical dynamical equilibrium and thermal equilibrium. From the simulations they have seen the nonlinear correlation between the SFR surface density $\Sigma$SFR and the total diffuse gas pressure at the midplane. They have argued that this correlation also applies to the starburst regime (the gas surface density $\Sigma$ $\sim$ $10^{2}$ $-$ $10^{4}$ $M_{\odot}pc^{-2}$). We plot their correlation in Fig.4 as comparison to the correlation of our sample. The slopes of the correlations are similar to each other. At a fixed $\Sigma$SFR, the thermal pressure in HII region in our local analogs is somewhat smaller than total midplane pressure in their simulations (by $\sim$ 0.3 dex). However, there are three main factors that we need to pay attention to when we do the comparison, due to the differences between the physical properties in this work and in their simulations. First, the local analogs are compact starbursts of ages $< 10^7$ years. They may not have had time to come into equilibrium yet. Second, we expect HII regions this young to be overpressured. Third, the thermal pressure is only a fraction of the total pressure, which also includes contributions from turbulence (a factor of 2 or more for Mach numbers $M>1$; \citet{elm00}), magnetic fields, and cosmic ray pressure.  The effects of these other sources of pressure will be to lower our observed thermal pressures below the total pressure that \citet{kim11} use, as seen in figure~4; while overpressure in the HII regions will have the opposite effect.  Overall, then, the correlation slope we have observed is broadly consistent with \citet{kim11}, and a modest offset of the correlation zero point (of either sign) appears physically plausible.

\section{Summary}
We have discussed the relation between the SFR surface density and the thermal pressure in HII regions for nearby (z $<$ 0.30) compact starbursts, with the sample of green peas, the nearby analogs of high-redshift Ly$\alpha$ emitters, and Lyman break analogs, the nearby analogs of high-redshift Lyman break galaxies.

1. We have measured the electron densities for a large sample of local analogs, which are 100 $\sim$ 700 cm$^{-3}$, comparable to the typical values for z $\sim$ 2 star-forming galaxies and larger than the typical values measured for SDSS star-forming galaxies. We have found that the electron temperature in HII regions for our sample is larger than the representative value of HII regions in z $\sim$ 0 star-forming galaxies, with the median value around 12000K. We have measured the size of the green pea galaxies in the high-resolution HST COS NUV images with GALFIT. We have found that the typical size of green peas galaxies is  $\sim$0.19 arcsec, and $\sim$0.7 kpc.

2. In our sample, green peas and Lyman break analogs have high $\Sigma$SFR up to 1.2 $M_{\odot}year^{-1}kpc^{-2}$ and high thermal pressure in HII region up to P/k$_B$ $\sim$10$^{7.2}$Kcm$^{-3}$, similar to the high pressures seen in local starburst which have massive outflows (e.g. M82). Large scale outflows are a necessary for the resonantly scattered Lyman-$\alpha$ photons to escape.

3.More importantly, we have found a correlation between SFR surface density and the thermal pressure in HII regions for the local analogs. This suggests a similar correlation in high-redshift Ly$\alpha$ emitters and Lyman break galaxies. 

4. The correlation, as well as the range of pressures, is consistent with the results from H$\alpha$ emitters at z $\sim$ 2.5 in \citet{shi15}.
 
\acknowledgments

Based in part on observations made with the NASA/ESA Hubble Space Telescope, obtained [from the Data Archive] at the Space Telescope Science Institute, which is operated by the Association of Universities for Research in Astronomy, Inc., under NASA contract NAS 5-26555. These observations are associated with program \#14201. Support for program \#14201 was provided by NASA through a grant from the Space Telescope Science Institute, which is operated by the Association of Universities for Research in Astronomy, Inc., under NASA contract NAS 5-26555.  In addition, we are grateful for partial financial support of this work from the US National Science Foundation through NSF grant AST-1518057. Tianxing Jiang thanks the LSSTC Data Science Fellowship Program, her time as a Fellow has benefited this work.
This work has made use of data from the Sloan Digital Sky Survey (SDSS). Funding for the SDSS and SDSS-II has been provided by the Alfred P. Sloan Foundation, the Participating Institutions, the National Science Foundation, the U.S. Department of Energy, the National Aeronautics and Space Administration, the Japanese Monbukagakusho, the Max Planck Society, and the Higher Education Funding Council for England. The SDSS Web Site is http://www.sdss.org/. The SDSS is managed by the Astrophysical Research Consortium for the Participating Institutions. The Participating Institutions are the American Museum of Natural History, Astrophysical Institute Potsdam, University of Basel, University of Cambridge, Case Western Reserve University, University of Chicago, Drexel University, Fermilab, the Institute for Advanced Study, the Japan Participation Group, Johns Hopkins University, the Joint Institute for Nuclear Astrophysics, the Kavli Institute for Particle Astrophysics and Cosmology, the Korean Scientist Group, the Chinese Academy of Sciences (LAMOST), Los Alamos National Laboratory, the Max-Planck-Institute for Astronomy (MPIA), the Max-Planck-Institute for Astrophysics (MPA), New Mexico State University, Ohio State University, University of Pittsburgh, University of Portsmouth, Princeton University, the United States Naval Observatory, and the University of Washington.

\clearpage

\begin{figure}
\includegraphics[scale=.37]{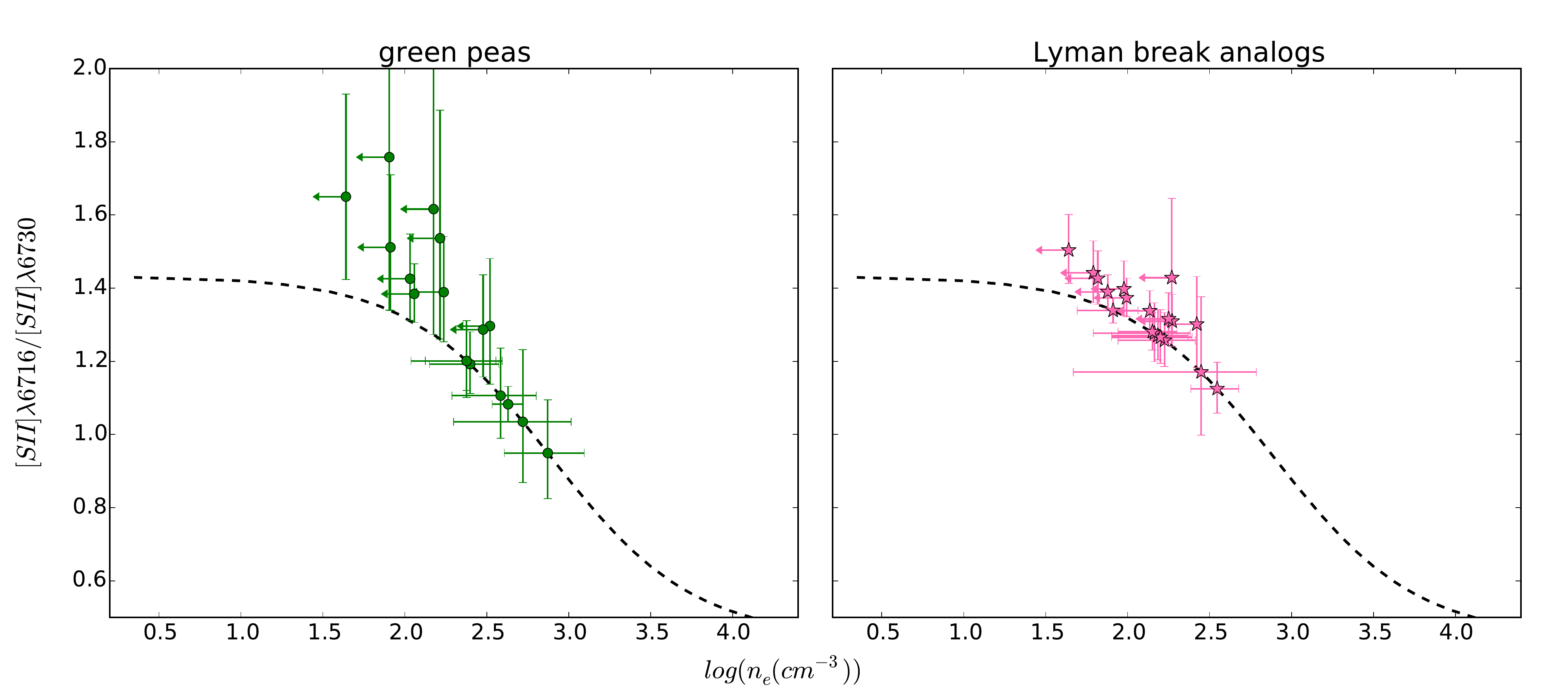}
\caption{SII line ratio vs electron density in HII region. The left panel shows green peas and the right panels show Lyman break analogs. The dashed line is a fit to the [SII] line ratio and electron density according to the IRAF routine ``temden.''}\label{fig1}
\end{figure}

\begin{figure}
\includegraphics[scale=0.75]{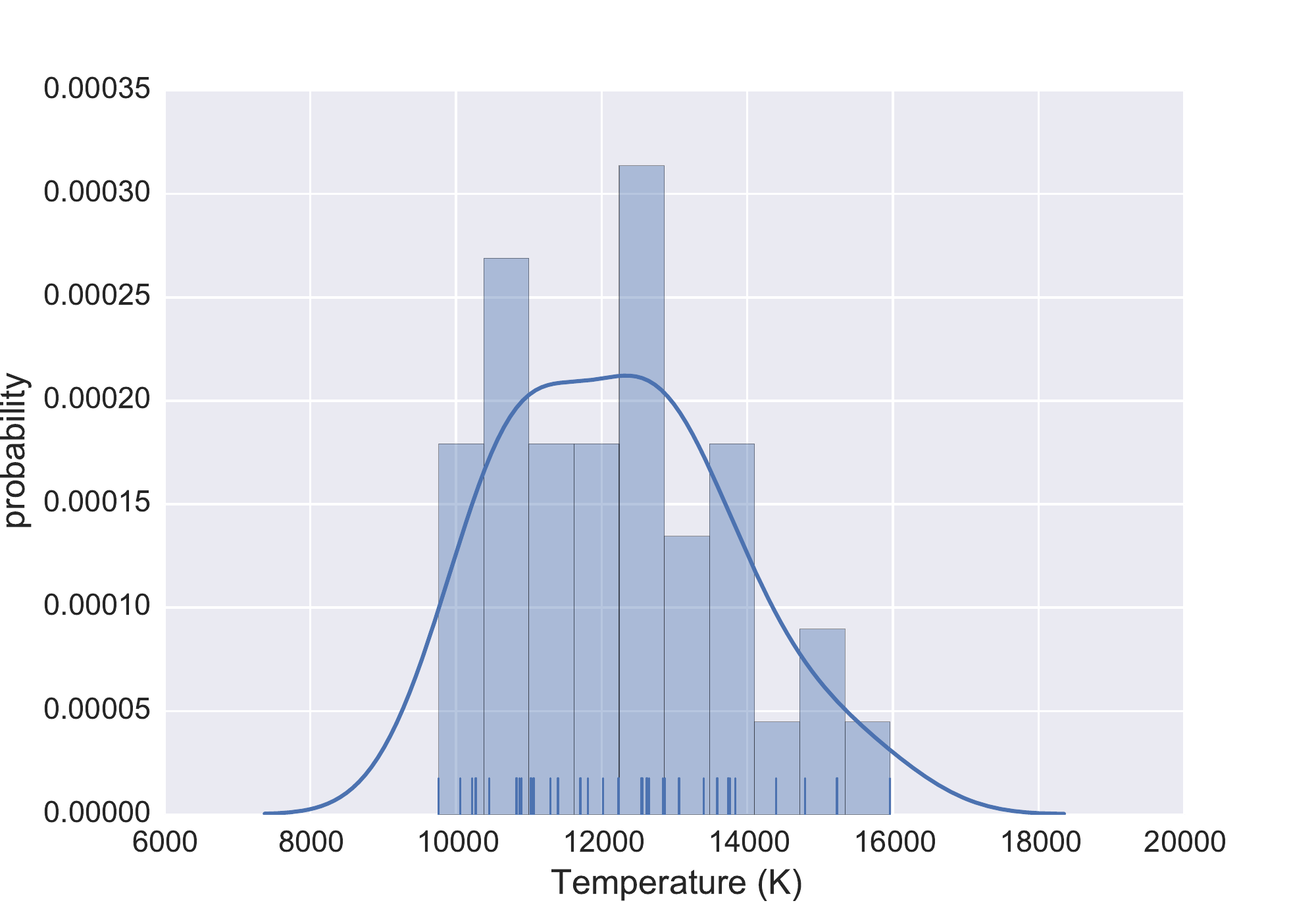}
\caption{Normalized histogram of the electron temperature measured for the ``parent sample''. The curve shows the kernel density estimate with the normal (Gaussian) kernel function. The kernel density estimate (KDE) is normalized such as the area under the KDE curve is equal to 1. The kernel density estimate is complementary to the histogram in presenting the distribution of a quantity. The numbers of galaxies in each bin, from left to right, are 4, 6, 4, 4, 7, 3, 4, 1, 2, 1, respectively.}\label{fig2}

\end{figure}

\begin{figure}
\includegraphics[scale=0.75]{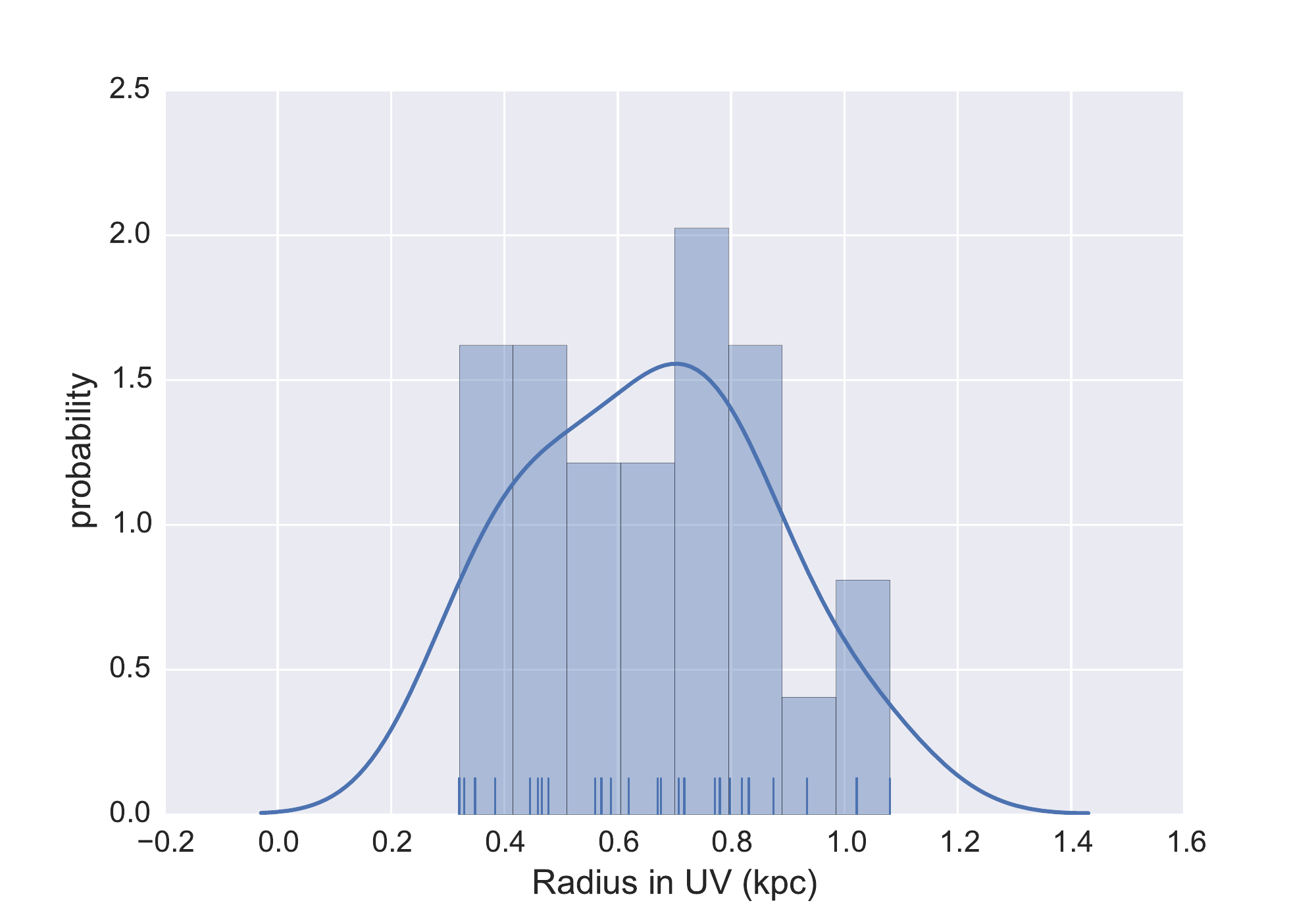}
\caption{Normalized histogram of the half-light radii of green peas in the ``parent sample''. The radii were measured in HST NUV images. The curve shows the kernel density estimate with the normal (Gaussian) kernel function. The numbers of galaxies in each bin, from left to right, are 4, 4, 3, 3, 4, 4, 1, 2, respectively.}\label{fig3}
\end{figure}

\begin{figure}
\includegraphics[scale=.58]{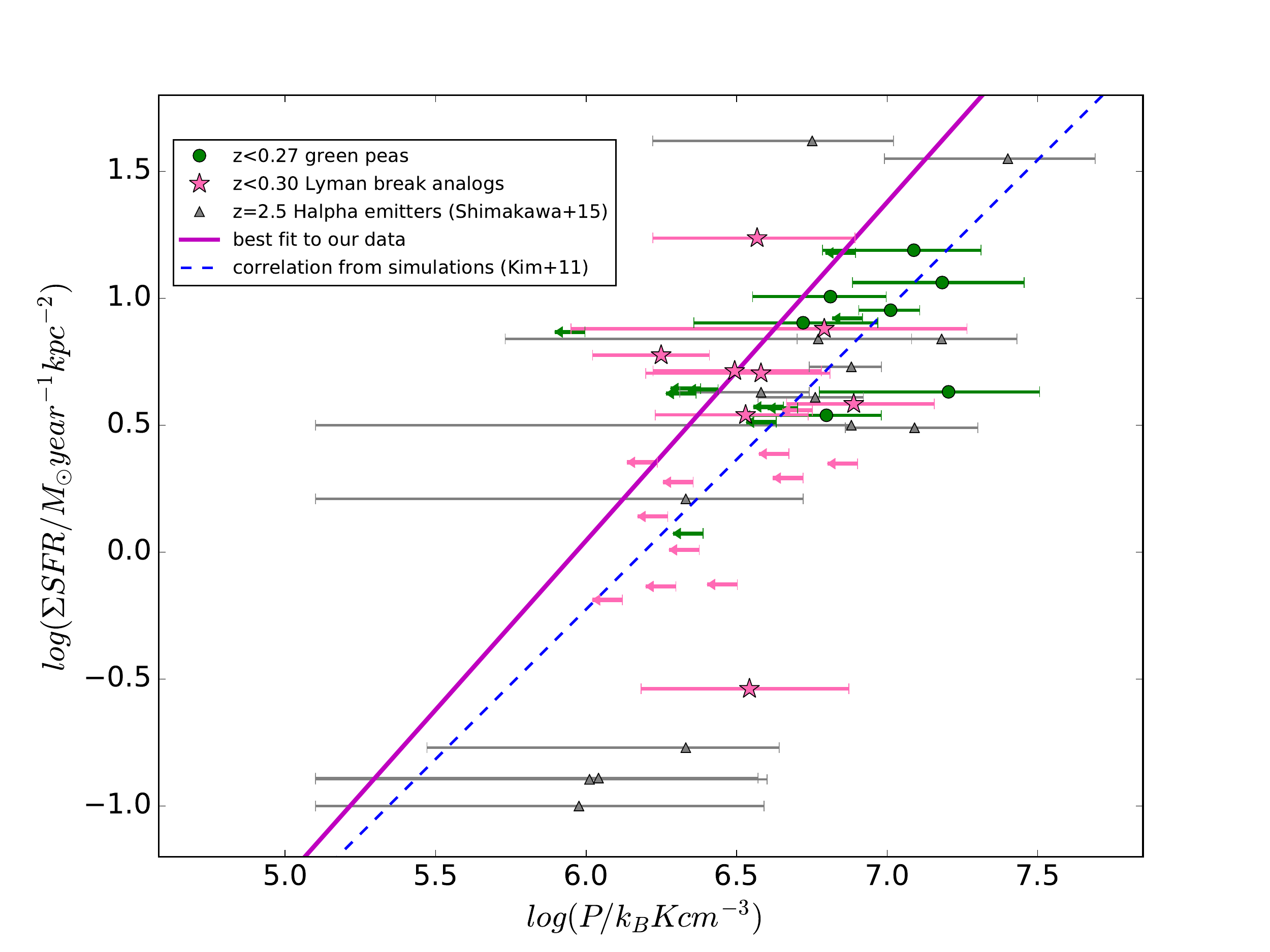}
\caption{SFR surface density vs pressure relation. The green filled circles and red stars (or green and red upper limits) are our sample. Note that in this figure, the thermal pressure of our sample is based on the electron density measurements that are listed in column 8, 9, 10 in Table 1, excluding the measurements labeled with e in column 8. Please refer to more details in the texts in Section 3.1 for the electron density measurements. The grey triangles are the H$\alpha$ emitters in \citet{shi15}. The best fit to our data is shown by the purple line. The correlation from the simulations in \citet{kim11} is the blue dashed line.}\label{fig5}
\end{figure}

\begin{figure}
\includegraphics[scale=.58]{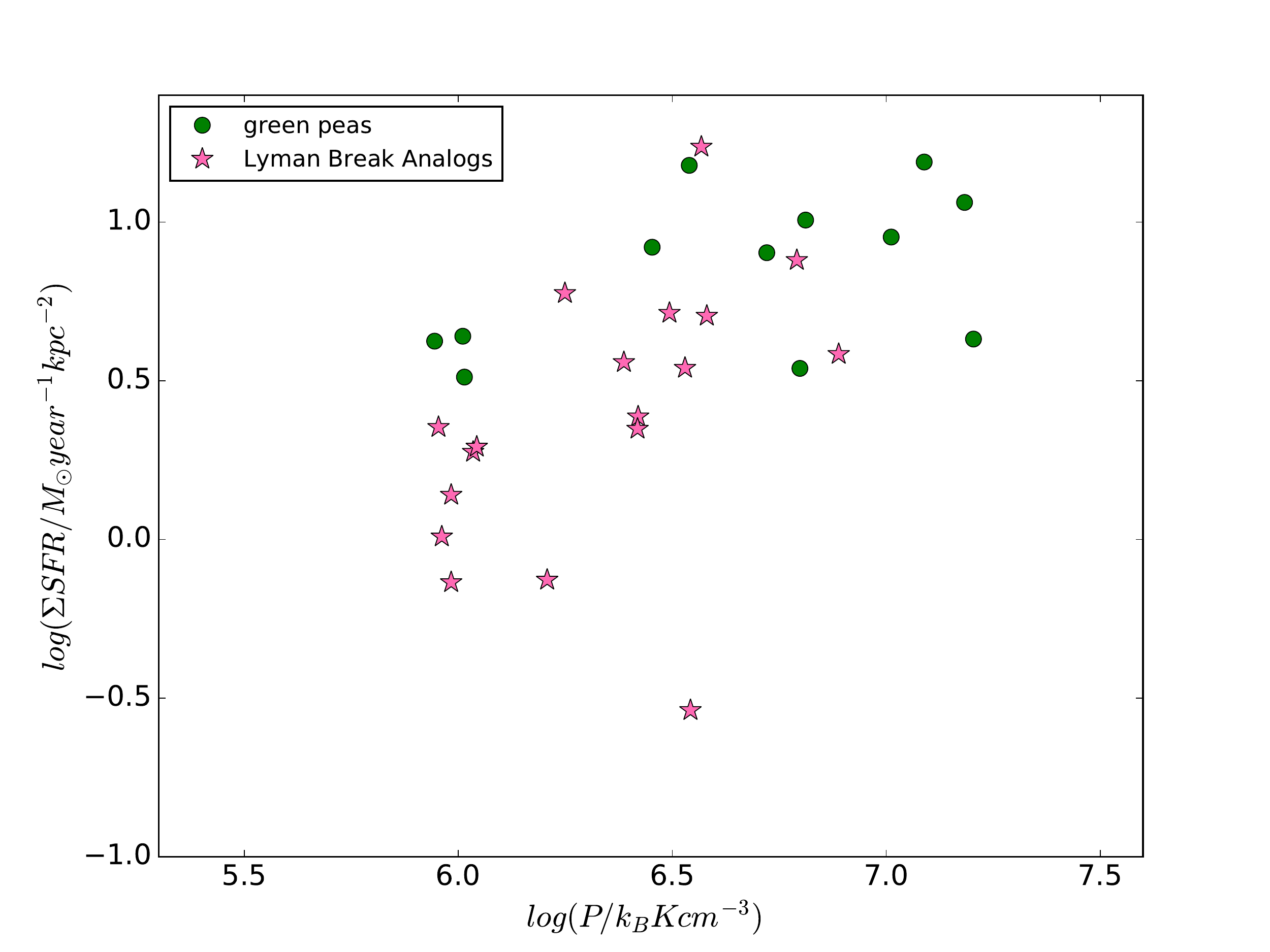}
\caption{SFR surface density vs thermal pressure (without error bars or upper limits) in HII regions for our sample, with green peas marked by green filled circles and Lyman break analogs marked by red stars. This figure is to show the data that are used in Spearman's rank correlation analysis. The thermal pressure is based on the electron density measurements that are listed in column 8 in Table 1. Details: For the objects with R $>$ 1.5, we could not get a reliable electron density measurement, so we excluded them for the Spearman's rank correlation analysis (not shown in this figure). For the objects that with R $\le$ 1.38, we measured the electron density from the line ratio (without considering the error bars). For objects with 1.38 $<$ R $<$ 1.5, we measured the electron density from a ratio of 1.38.}\label{fig4}
\end{figure}

\begin{figure}
\includegraphics[scale = .42]{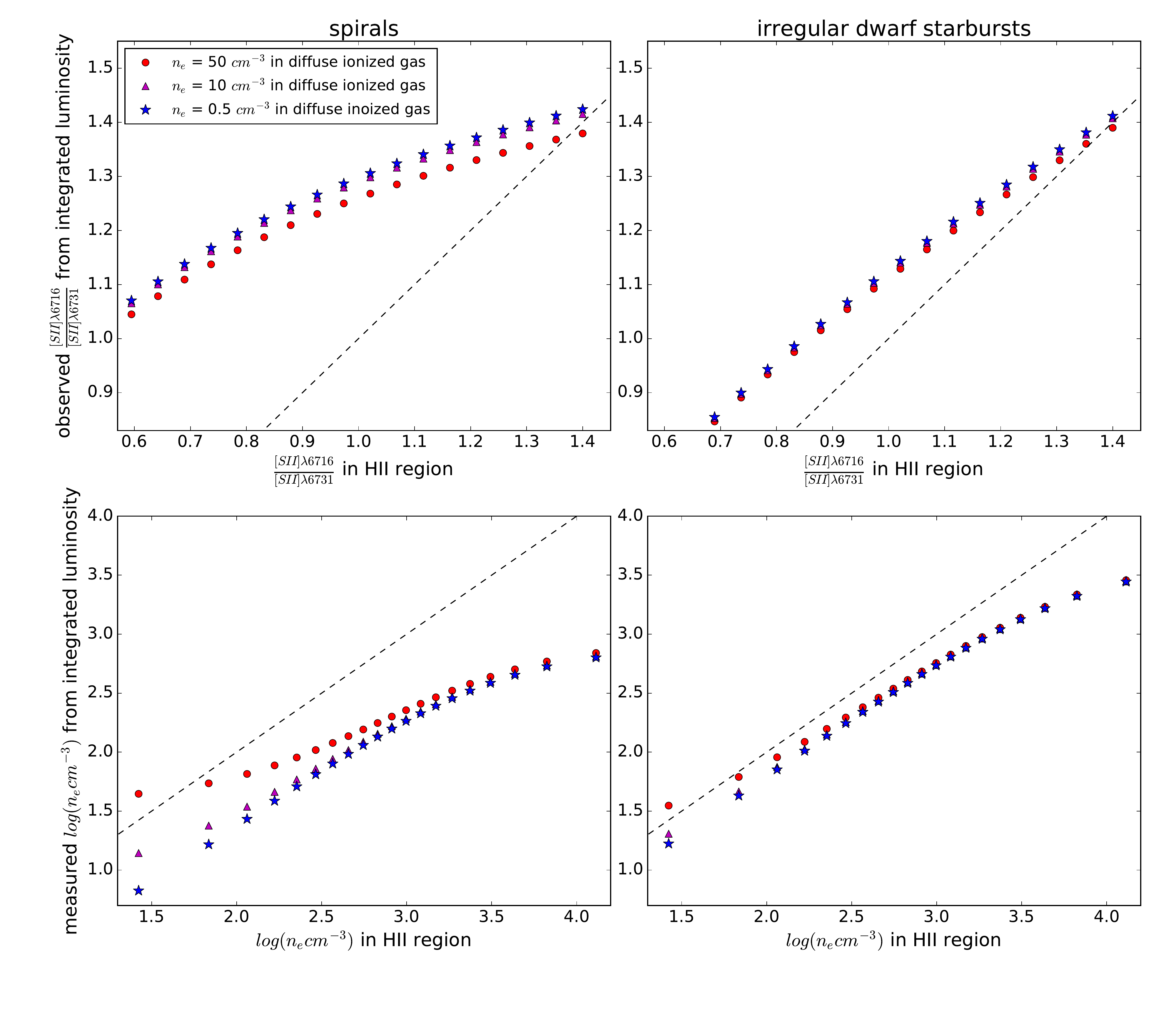}
\caption{The observed [SII] emission from galaxies is a superposition of [SII] within HII regions and [SII] from diffuse gas outside HII regions. In this figure, we explore the implications of this superposition for our study. {\it Upper panels:}\/ The observed [SII] line ratio from integrated luminosity vs the [SII] line ratio from HII region. {\it Lower panels:}\/ The electron density measured from integrated luminosity vs the electron density in HII region. The left panels are for physical conditions representative of spirals, and the right panels are for conditions representative of irregular dwarf starbursts (see text for details). The dashed line in each panel shows the location of x = y. The three symbols show three different assumptions of the electron density in the diffuse ionized gas, with red filled circles marking 50 cm$^{-3}$, purple triangles marking 10 cm$^{-3}$, and blue stars marking 0.5 cm$^{-3}$.  In general, the inferred electron density $n_e$ is between the true electron density in HII regions and the (generally lower) electron density in the diffuse gas.  The magnitude of the effect depends on assumed physical parameters, but is generally 0.2--0.4 dex for our dwarf starburst models.}
\end{figure}

\clearpage

\begin{deluxetable}{ccccccccccc}

\tabletypesize{\scriptsize}

\tablecaption{Properties of the Green Peas and Lyman Break Analogs}

\tablenum{1}

\tablehead{\colhead{ID} & \colhead{RA\tablenotemark{a}} & \colhead{Dec.\tablenotemark{a}} & \colhead{z\tablenotemark{b}} & \colhead{R$_{e}$\tablenotemark{c}} & \colhead{SFR\tablenotemark{d}} & \colhead{n$_{e}$} & \colhead{n$_{e}$ u68\tablenotemark{f}} & \colhead{n$_{e}$ l68\tablenotemark{g}} & \colhead{T$_{e}$} \\ 
\colhead{} & \colhead{(J2000)} & \colhead{(J2000)} & \colhead{} & \colhead{(kpc)}  & \colhead{(M$_{\odot}$ yr$^{-1}$)} & \colhead{(cm$^{-3}$)} & \colhead{(cm$^{-3}$)} & \colhead{(cm$^{-3}$)} & \colhead{(10$^4$ K)} }

\startdata
GP01 & 03:03:21.41 & -07:59:23.25 & 0.164 & 0.56  & 8.41 & 525 & 1036 & 198 & 1.52 \\
GP02 & 12:44:23.37 & 02:15:40.43 & 0.239 & 1.02  & 22.69 & 250 & 375 & 141 & 1.25 \\
GP03 & 10:53:30.82 & 52:37:52.87 & 0.253 & 0.62 & 17.76 & \nodata   & 44 & \nodata  & 1.08 \\
GP04 & 14:24:05.73 & 42:16:46.29 & 0.185 & 0.48 & 14.56 & 238 & 360 & 133 & 1.36 \\
GP05 & 12:19:03.98 & 15:26:08.51 & 0.196 & 0.33 & 10.57 & 384 & 633 & 194 & 1.60 \\
GP06 & 11:37:22.14 & 35:24:26.69 & 0.194 & 0.72 & 14.16 & 44\tablenotemark{e} & 114 & \nodata  & 1.17 \\
GP07 & 09:11:13.34 & 18:31:08.17 & 0.262 & 0.57 & 17.07 & 124\tablenotemark{e} & 331 & \nodata  & 1.14 \\
GP08 & 08:15:52.00 & 21:56:23.65 & 0.141 & 0.35 & 3.36 & \nodata  & 82 & \nodata  & 1.44 \\
GP09 & 08:22:47.66 & 22:41:44.08 & 0.216 & 0.68 & 25.79 & 427 & 523 & 341 & 1.20 \\
GP10 & 03:39:47.79 & -07:25:41.28 & 0.261 & 0.87 & 20.26 & 44 & 108 & \nodata  & 1.01 \\
GP11 & 22:37:35.06 & 13:36:47.02 & 0.294 & 1.08 & 23.76 & 44 & 173 & \nodata  & 1.18 \\
GP12 & 14:54:35.59 & 45:28:56.24 & 0.269 & 0.44 & 14.37 & 746 & 1245 & 405 & 1.02 \\
GP13 & 14:40:09.94 & 46:19:36.95 & 0.301 & 0.72 & 25.91 & 238 & 393 & 109 & 1.10 \\
GP14 & 07:51:57.78 & 16:38:13.24 & 0.265 & 0.80 & 4.73 & \nodata  & 80 & \nodata  &  1.29 \\
GP15 & 10:09:19.00 & 29:16:21.50 & 0.222 & 0.46 & 4.88 & \nodata  & 164 & \nodata  & 1.48 \\
GP16 & 12:05:00.67 & 26:20:47.74 & 0.343 & 0.83 & 16.23 & \nodata  & 150 & \nodata  & 1.22 \\
GP17 & 13:39:28.30 & 15:16:42.13 & 0.192 & 0.38 & 13.97 & 135 & 301 & \nodata  & 1.28 \\
LBA01 & 00:55:27.46 & 00:21:48.71 & 0.167 & 0.77   & 4.41 & 352 & 475 & 243 & 1.10\tablenotemark{h}   \\
LBA02 & 01:50:28.41 & 13:08:58.40 & 0.147 & 1.83   & 14.69 & 44\tablenotemark{e} & 76 & \nodata  & 1.03 \\
LBA03 & 02:03:56.91 & -08:07:58.51 & 0.189 & 1.61   & 9.52 & 50\tablenotemark{e} & 99 & \nodata  & 1.09 \\
LBA04 & 03:28:45.99 & 01:11:50.85 & 0.142 & 1.82   & 4.79 & 83\tablenotemark{e} & 137 & \nodata  & 0.98 \\
LBA05 & 03:57:34.00 & -05:37:19.70 & 0.204 & 1.09  & 8.34 & 111\tablenotemark{e} & 187 & \nodata  & 1.10\tablenotemark{h}  \\
LBA06 & 04:02:08.87 & -05:06:42.06 & 0.139 & 1.42   & 2.53 & \nodata  & 44 & \nodata  & 1.10\tablenotemark{h}  \\
LBA07 & 08:20:01.72 & 50:50:39.16 & 0.217 & 1.52  & 15.57 & 153 & 234 & 80 & 1.11 \\
LBA08 & 08:25:50.95 & 41:17:10.30 & 0.156 & 1.56   & 6.52 & 44\tablenotemark{e} & 62 & \nodata  & 1.10\tablenotemark{h}  \\
LBA09 & 08:38:03.73 & 44:59:00.28 & 0.143 & 0.92   & 4.01 & 104\tablenotemark{e} & 178 & \nodata  & 1.26 \\
LBA10 & 09:23:36.46 & 54:48:39.25 & 0.222 & 0.48   & 7.71 & 168 & 259 & 87 & 1.10\tablenotemark{h}  \\
LBA11 & 09:26:00.41 & 44:27:36.13 & 0.181 & 1.09  & 11.71 & 146 & 241 & 62 & 1.31 \\
LBA12 & 09:38:13.50 & 54:28:25.09 & 0.102 & 0.92   & 9.85 & 82 & 116 & 49 & 1.09 \\
LBA13 & 10:26:13.97 & 48:44:58.94 & 0.160 & 1.99  & 7.83 & 44\tablenotemark{e} & 95 & \nodata  & 1.05 \\
LBA14 & 12:48:19.75 & 66:21:42.68 & 0.260 & 1.9   & 15.67 & 119\tablenotemark{e} & 264 & \nodata  & 1.10\tablenotemark{h} \\
LBA15 & 13:53:55.90 & 66:48:00.59 & 0.198 & 3.57   & 18.10 & 44\tablenotemark{e} & 66 & \nodata  & 1.10\tablenotemark{h}  \\
LBA16 & 14:34:17.16 & 02:07:42.58 & 0.180 & 4.6  & 11.87 & 159 & 247 & 80 &  1.10\tablenotemark{h}\\
LBA17 & 21:45:00.26 & 01:11:57.58 & 0.204 & 1.16 & 13.54 & 142 & 200 & 87 &  1.10\tablenotemark{h}\\
LBA18 & 23:25:39.23 & 00:45:07.25 & 0.277 & 0.81 & 9.70 & 281 & 610 & 47 &  1.10\tablenotemark{h}\\
LBA19 & 23:53:47.69 & 00:54:02.08 & 0.223 & 1.31  & 6.53 & 44 & 186 &  \nodata & 1.26 \\
\enddata
\tablecomments{}
\tablenotetext{a}{For green peas, the Ra and Dec. are from \citet{car09}. For LBAs, the Ra and Dec. are from \citet{ove09}.}
\tablenotetext{b}{For green peas, the redshift is based on H$\alpha$ emission line in SDSS DR12. For LBAs, the redshift is from \citet{ove09}.}
\tablenotetext{c}{Half-light radius. For green peas, this is the half-light radius measured in HST NUV images. For LBAs, this is from \citet{ove09} measured in HST optical images.}
\tablenotetext{d}{The star formation rate is measured by us from the MPA H$\alpha$ luminosities.}
\tablenotetext{e}{These values are only used in Fig.5 for Spearman's rank correlation analysis but not used in Fig.4. Please refer to the caption of Fig.5 or Section 4 for the details. The other values in this column are used in both Fig.4 and Fig.5.}
\tablenotetext{f}{The upper 1$\sigma$ bound is measured based on the lower 1$\sigma$ bound of the [SII] $\lambda$6716 / $\lambda$6731 ratio.}
\tablenotetext{g}{The lower 1$\sigma$ bound is measured based on the upper 1$\sigma$ bound of the [SII] $\lambda$6716 / $\lambda$6731 ratio.}
\tablenotetext{h}{The value of 11000.0 K is assumed as the electron temperature of the objects for which the temperature can not be measured from [OIII] lines.}

\end{deluxetable}

\clearpage

\end{document}